\documentstyle[12pt,epsf]{article}
\begin{document}
\def\gsim{{~\raise.15em\hbox{$>$}\kern-.85em \lower.35em\hbox{$\sim$}~}}
\def\lsim{{~\raise.15em\hbox{$<$}\kern-.85em \lower.35em\hbox{$\sim$}~}}
\begin{titlepage}
\vfill

\hskip 3.5in ISU-HET-97-6

\hskip 3.5in October, 1997

\vskip 1.2in
\begin{center}
{\large \bf Long Distance Contribution to $K_L \rightarrow \ell^+ \ell^-$ }\\

\vspace{1 in}
{\bf  G.~Valencia}\\
{\it           Department of Physics and Astronomy\\
               Iowa State University\\
               Ames IA 50011}\\
{\it e-mail:valencia@iastate.edu}\\
\vspace{1 in}
\end{center}
\begin{abstract}

We revisit the calculation of the long distance contribution to 
$K_L \rightarrow \mu^+ \mu^-$. We discuss this process within the 
framework of chiral perturbation theory, and also using simple 
models for the $K_L \gamma^* \gamma^*$ vertex. We argue that it 
is unlikely that this mode can be used to extract information on 
short distance parameters. The process $K_L \rightarrow e^+ e^-$ 
is also long-distance dominated and we find that $B(K_L \rightarrow 
e^+ e^-) \approx 9 \times 10^{-12}$. 

\end{abstract}
\end{titlepage} 
\clearpage

\section{Introduction}
 
The rare decay mode $K_L \rightarrow \mu^+ \mu^-$ has been the 
subject of  many discussions
\cite{gailee,quijak,berg,belgeng,rrw,liva}. 
The current 
interest in this mode arises because experiment E871 at Brookhaven 
National Laboratory expects to collect a sample of about $10^4$ 
events \cite{rrw}. Theoretically, the interest in this decay 
mode focuses almost exclusively in the possibility of extracting 
constraints on $\rho$ in the Wolfenstein parameterization of the 
CKM matrix.\footnote{If the muon polarization is measured 
the decay is also interesting for $CP$ violation \cite{herczeg}.}   

The measured rate can be attributed to the sum of an absorptive 
part arising from a two-photon intermediate state, and a real 
part arising from the coherent sum of the dispersive contribution of the
two photon intermediate state and a short distance contribution that
depends on $\rho$. Therefore, to extract information on $\rho$, 
one must be able to calculate reliably the dispersive contribution 
associated with the two photon intermediate state.

The short distance contribution to this process has been  
known for quite some time \cite{gailee,inami}. A recent review that 
presents all the relevant details of the status of this calculation 
is Ref.~\cite{bufle}. For our purposes it will suffice to quote an 
approximate result for $m_t = 175$~GeV and $V_{cb}=0.040$ \cite{burasb}:
\begin{equation}
B(K_L \rightarrow \mu^+ \mu^-)_{SD} = 7.9\times 10^{-10}  
(\rho_0 -\rho)^2
\label{kmmbu}
\end{equation}
$\rho_0 \sim 1.2$ is a measure of the charm-quark contribution 
to this decay.

In order to discuss the two-photon contribution, it is convenient to 
normalize the decay rate to the two-photon mode. Using the notation 
$\beta_\ell \equiv \sqrt{1-4m_\ell^2/ M_K^2}$ we have:
\begin{equation}
{B(K_L \rightarrow \ell^+ \ell^-)\over B(K_L \rightarrow \gamma \gamma)}
= 2 \biggl({m_\ell \over M_K}\biggr)^2\biggl({\alpha_{em}\over \pi}\biggr)^2
\beta_\ell\biggl[({\rm Re}A)^2+({\rm Im}A)^2\biggr],
\label{notation}
\end{equation}
The absorptive contribution to the rate is determined uniquely, it 
corresponds to:
\begin{equation}
{\rm Im}A={\pi \over 2\beta_\ell}\log\biggl({1+\beta_\ell 
\over 1-\beta_\ell}\biggr).
\label{absorptive}
\end{equation}
If we use the branching ratio quoted in the Particle Data Book
\cite{pdb}:
\begin{equation}
B(K_L \rightarrow \gamma \gamma) = (5.92 \pm 0.15)\times 10^{-4}
\label{brggdata}
\end{equation}
we find that the absorptive part of the two-photon contribution to 
$K_L \rightarrow \mu^+ \mu^-$ is:
\begin{equation}
B(K_L \rightarrow \mu^+ \mu^-)_{\rm ABS} = (7.1 \pm 0.2)\times 10^{-9}.
\label{brabs}
\end{equation}
At the same time, the rate for $K_L \rightarrow \mu^+ \mu^-$ has been 
measured to be:
\begin{equation}
B(K_L \rightarrow \mu^+ \mu^-) = \cases{
(7.2 \pm 0.5) \times 10^{-9} & PDB \cite{pdb} \cr
(6.86 \pm 0.37)\times 10^{-9} & AGS-791 \cite{heinson} \cr
(7.9 \pm 0.7)\times 10^{-9} & KEK 137 \cite{mumu137}\cr} 
\label{brmmdata}
\end{equation}
If we use, for definiteness, the Particle Data Book average, 
we find that the absorptive part almost completely saturates 
the experimental rate and that there is room for a small real 
part in the amplitude (arising from the coherent sum of the 
short distance and two-photon dispersive contributions). Taking 
into account that the total rate must be greater than or equal 
to the absorptive contribution, we can write:
\begin{equation}
{\rm Re}A = 0 \pm 1.2
\label{roomfora}
\end{equation}

There exist several models in the literature that allow one to estimate
the size of the long distance two-photon contribution. However, these 
models do not provide a reliable estimate of the uncertainty in their 
predictions. Nevertheless, a recent paper \cite{ampoisi} claims to 
obtain a reliable estimate of the magnitude and uncertainty of the 
two-photon contribution and from this a lower bound on $\rho$.

In this paper we revisit the issue of the two-photon long distance
contribution to $K_L \rightarrow \mu^+ \mu^-$. We first use the framework 
of chiral perturbation theory to parameterize the 
dispersive part of the two-photon contribution in terms 
of one unknown combination of constants. We argue that this combination cannot 
be measured in any other process and, therefore, that 
it is not possible to remove
the uncertainty. We then turn our attention to a simple pole model for 
the $K_L\gamma^*\gamma^*$ vertex that we use to discuss some 
of the model calculations that have appeared in the literature and to 
offer an estimate for the uncertainty of the result. Finally, we 
discuss a slightly more complicated model that illustrates why a 
more precise measurement of the decay parameters in  
$K_L \rightarrow \ell^+ \ell^- \gamma$ will not reduce the uncertainty 
in the dispersive two-photon contribution to $K_L \rightarrow \ell^+ 
\ell^-$. We conclude that it is not possible at present to extract 
information on the short distance parameter $\rho$ from the 
measurement of $K_L \rightarrow \mu^+ \mu^-$.

\section{Chiral Perturbation Theory}

To calculate the dispersive part of the two-photon intermediate 
state it is not sufficient to know the $K_L \rightarrow \gamma 
\gamma$ amplitude on shell, so we parameterize the $K_L \gamma^*\gamma^*$ 
vertex in the following way:
\begin{equation}
\Gamma^{\mu\nu}(q_1,q_2)=
2 {\alpha_{em} \over \pi} G_8 f_\pi F(q_1^2,q_2^2)
\epsilon_{\alpha\beta\mu\nu}q_1^\alpha q_2^\beta .
\label{akgg}
\end{equation}
The overall constant, with
$G_8 \approx 5.1 G_F /\sqrt{2}|V_{ud}V^*_{us}|$, 
has been chosen for convenience \cite{liva}. With this normalization, 
the measured rate for $K_L \rightarrow \gamma \gamma$, Eq.~\ref{brggdata}, 
implies the value 
\begin{equation}
F_1 \equiv F(0,0)= 0.90 \pm 0.01
\label{measuref}
\end{equation} 
The behavior of the vertex $\Gamma^{\mu\nu}(q_1,q_2)$ (Eq.~\ref{akgg}) 
as a function of $q_1^2$ and $q_2^2$ is 
described by the function $F(q_1^2,q_2^2)$. 
This function is symmetric under the interchange 
$q_1 \leftrightarrow q_2$ due to Bose symmetry. 
For energies that are small compared to the scale of chiral symmetry 
breaking, $\Lambda \sim 1$~GeV, we find it convenient to write 
this form factor as:
\begin{equation}
F(q_1^2,q_2^2) = F(0,0)\biggl( 1 + \alpha {(q_1^2 +q_2^2)\over \Lambda^2} 
+ \beta {(q_1^4 +q_2^4)\over \Lambda^4} + 
\beta^\prime {q_1^2 q_2^2 \over \Lambda^4} + \cdots
\biggr)
\label{expandf}
\end{equation}
It should be possible to extract the first few parameters 
in this expansion from measurements in the modes $K_L \rightarrow 
\ell^+ \ell^- \gamma$. In fact, a study of $K_L \rightarrow 
e^+ e^-\gamma$ has already found \cite{keeg} 
\begin{equation}
\alpha = 3.15 \pm 0.40.
\label{alpha}
\end{equation}
Measurements of $K_L \rightarrow \mu^+ \mu^- \gamma$ are 
consistent with this number at the two standard deviation level 
\cite{kmumug}. It should be possible to improve the precision of this 
number, as well as to measure the parameters $\beta$ and $\beta^\prime$ 
in the current generation of experiments. In this regard, it is 
important to emphasize that it would be valuable to have a direct 
fit to these parameters from the experiments, instead of the 
fit to the parameter $\alpha_{K^*}$ that is currently 
performed. This was already suggested in References~\cite{liva} 
and~\cite{ampoisi}.  

The measured values of $F_1$ and $\alpha$, justify the way we wrote 
the low energy expansion in Eq.~\ref{expandf}. However, in chiral 
perturbation theory one finds that $F(q_1^2,q_2^2)$ vanishes at 
order $p^4$ \cite{lin}. The leading contributions appear at order 
$p^6$, and this implies that, formally, $F_1$ and $\alpha$ are of the 
same order. To be truly consistent with $\chi$PT we should write 
$M_K^2/\Lambda^2$ instead of one in the first term of Eq.~\ref{expandf}. 
As we will see, this complicates the discussion without  
altering the conclusions, so we leave Eq.~\ref{expandf} as it 
stands.

Let us first evaluate the one-loop, two-photon intermediate state, 
contribution to $K_L \rightarrow \ell^+ \ell^-$. To leading order 
in $\chi$PT, the contribution of this loop is obtained 
by using the vertex of Eq.~\ref{akgg} with the form factor 
of Eq.~\ref{expandf} truncated at the term linear in $q_i^2$, and
attaching the photons to the leptons with the usual rules of QED. The 
resulting one-loop diagram is divergent and the calculation of the physical 
process requires local counter-terms.

The local counter-terms are written as a chiral Lagrangian 
using standard notation \cite{dobook}. The pion and kaon fields are 
identified with the Goldstone bosons of the spontaneously broken  
$SU(3)_L\times SU(3)_R$ chiral symmetry and are incorporated via the 
matrix:
\begin{eqnarray}
\phi &=& {1 \over \sqrt{2}}
\left( \begin{array}{ccc}
\pi^0 /\sqrt{2}+\eta /\sqrt{6} & \pi^+ & K^+ \\
\pi^- &-\pi^0 /\sqrt{2}+\eta /\sqrt{6} & K^0 \\
K^- & \overline{K^0} & -2 \eta /\sqrt{6}
       \end{array} \right)
\label{pions}
\end{eqnarray}
into the matrix $\Sigma = \exp(i2\phi/f_\pi)$, which transforms 
as $\Sigma \rightarrow L \Sigma R^\dagger$ under 
$SU(3)_L\times SU(3)_R$. Interactions with external fields are  
incorporated by the use of suitable covariant derivatives, and 
by including terms with the fields $F_{\mu\nu}^{L,R}$ 
that transform under $SU(3)_L\times SU(3)_R$ as 
$L F_{\mu\nu}^LL^\dagger$ and $R F_{\mu\nu}^R R^\dagger$, respectively.
For electromagnetic interactions, the only external fields of 
interest are photons, so 
$F_{\mu\nu}^L=F_{\mu\nu}^R= e Q F_{\mu\nu}$. The matrix 
$Q$ is the diagonal matrix $(2/3,-1/3,-1/3)$.

To construct the local counter-terms 
for the processes $K_L \rightarrow \ell^+ \ell^-$ 
and $K \rightarrow \pi \ell^+ \ell^-$ we use the 
following information: a) they will have two factors 
of $Q$, originating from two photon fields that are  
integrated out; b) they will contain the lepton fields in a 
parity odd bilinear $\overline{\ell}\gamma_5 \ell$ or  
$\overline{\ell}\gamma_\mu \gamma_5\ell$ to conserve $CP$ 
in $K_L \rightarrow \ell^+ \ell^-$;   
c) we assume octet dominance and $CPS$ symmetry for the 
weak interactions; d) we use the result  $[Q,\lambda_6]=0$. All 
this information results in the lowest dimension counter-terms:
\begin{eqnarray}
{\cal L}& = & i F_1 G_8 \biggl({\alpha_{em} \over \pi}\biggr)^2f^2_\pi
 \overline{\ell}\gamma_\mu \gamma_5 \ell \biggl[
h_1 \mbox{Tr}\biggl(\lambda_6 Q^2 \bigl(\Sigma \partial^\mu 
\Sigma^\dagger-\partial^\mu \Sigma \Sigma^\dagger \bigr)\biggr)
\nonumber \\
&+& h_2\mbox{Tr}\biggl(\lambda_6 Q \bigl(\Sigma Q\partial^\mu 
\Sigma^\dagger -\partial^\mu\Sigma Q \Sigma^\dagger \bigr)\biggr)
\nonumber \\
&+&h_3 \mbox{Tr}\biggl(\lambda_6 \bigl(\Sigma Q^2 \partial^\mu 
\Sigma^\dagger -\partial^\mu \Sigma Q^2 \Sigma^\dagger \bigr) \biggr)
\biggr]
\label{lwct}
\end{eqnarray}
where we have included the overall factor $F_1$ from 
Eq.~\ref{measuref} for convenience. Notice that we have not listed 
counter-terms with no derivatives, such as 
$\mbox{Tr}\bigl(\lambda_6\Sigma Q^2 \Sigma^\dagger\bigr) 
\overline{\ell}\gamma_5 \ell$, because they do not respect $CPS$ symmetry.

Since these are the lowest dimension counter-terms that can be 
constructed for this process, we expect them to dominate based 
on power counting. As we will see after the explicit evaluation of the 
one-loop amplitude, however, the non-analytic terms that arise 
are the dominant ones. 

It is straightforward to carry out the one-loop calculation  
using dimensional regularization and standard techniques. It is 
convenient to express the result in terms of a one-dimensional 
integral over a Feynman parameter, and to carry out that last 
integration numerically. We find:
\begin{eqnarray}
{\rm Re}A &=&-{8 \over 9\sqrt{2}}(h_1+h_2+h_3)
+{3\over 2}\biggl[{1\over \hat\epsilon}-{2\over 3}
-\log\bigl({m^2_\ell \over \mu^2}\bigr) \biggr] 
+{7\over 2} \nonumber \\
&+& 2\xi^2+2\xi^4+\xi^2\log(4\xi^2)+2\xi^4\log(4\xi^2)\nonumber \\
&+& \int_0^1 dx\biggl[2{\lambda_+ + \lambda_- \over \lambda_+ - \lambda_-}
\bigl(2x\xi^4+\xi^2-2x+2\bigr)\log\biggl({\lambda_-\over\lambda_+}\biggr)
\biggr] \nonumber \\
&+& 
{\alpha \over \Lambda^2}\biggl[\biggl({1\over \hat\epsilon}-
\log\bigl({m^2_\ell \over \mu^2}\bigr) \biggr)\biggl({10\over 3}
m_\ell^2 -{1\over 3}M_K^2\biggr)+{26\over 9}m_\ell^2
-{2\over 9}M_K^2\biggr]
\label{oneloop}
\end{eqnarray}
We have used the notation $\xi^2=M^2_K/(4m_\ell^2)$, 
$\lambda_\pm= \xi \sqrt{x} \pm \sqrt{\xi^2 x+1-x}$, and 
$1/\hat\epsilon=2/(4-n)+\log 4\pi-\gamma$.

The terms generated by the loop {\it appear} to be of the same order 
as the tree-level constant $h_1+h_2+h_3$ in Eq.~\ref{oneloop}. This 
is due to the factor $F_1$ that we introduced as a normalization in 
Eq.~\ref{lwct}. However, as discussed earlier, $F_1$ itself is 
formally of order $p^2$. Therefore, the constant $h_1+h_2+h_3$ is
formally the leading contribution to the amplitude. For the same reason,  
the terms proportional to $\alpha$ are formally of the same order as the
other terms from the loop. Numerically, however, the 
finite terms proportional to $\alpha$ are smaller than the rest, and 
the finite loop-induced terms are as large as, or 
larger than, the tree-level constant.

Strictly speaking, the divergences that appear in Eq.~\ref{oneloop} 
should be renormalized by counter-terms with dimension 
higher than those of Eq.~\ref{lwct}. In view of our discussion above, 
however, we will combine these higher order renormalized counter-terms 
with the $h_i$ in the expression,
\begin{equation}
h(\mu)= h_1+h_2+h_3 + H^\prime -{9 \sqrt{2}\over 8}
{1\over \hat\epsilon}\biggl[{3\over 2}
-{\alpha\over 3\Lambda^2}(10m_\ell^2-M_K^2)\biggr]
\label{hren}
\end{equation}
where $H^\prime$ stands for the contributions from the counter-terms of 
dimension higher than those in Eq.~\ref{lwct} that we have not 
included. From the structure of Eq.~\ref{hren} we expect that 
$H^\prime$ will consist of the product of the coupling constants of 
the higher dimension counter-terms and factors of the lepton and kaon 
mass. The only purpose of $H^\prime$ 
is to renormalize Eq.~\ref{oneloop}, and once this 
is accomplished, we may treat $H^\prime$ as a higher order correction to 
$h_1+h_2+h_3$. Our prescription is, thus, to keep only the leading order 
counter-terms and the leading one-loop terms defined by the 
subtraction scheme embodied in Eq.~\ref{hren}.

The expression, Eq.~\ref{oneloop}, can be compared to the one obtained in 
Ref.~\cite{simipi} for the processes $\pi^0, \eta \rightarrow \ell^+
\ell^-$. Taking into account the different overall normalizations, 
and the different counter-terms, our expression agrees with that 
of Ref.~\cite{simipi} in the limit $\alpha=0$.

It is convenient to extract the leading behavior of Eq.~\ref{oneloop} for 
small lepton mass by re-writing it in the form:
\begin{eqnarray}
{\rm Re}A &=& -{8 \over 9\sqrt{2}}h(\mu)-
\log\bigl({m^2_\ell \over \mu^2}\bigr)
\biggl[{3\over 2}+{\alpha\over 3 \Lambda^2}(10m_\ell^2-M_K^2)
\biggr]-
\biggl(\log\bigl({m_\ell \over M_K}\bigr)\biggr)^2
\nonumber \\
&+& {5\over 2} + {\alpha\over 9\Lambda^2}(26m_\ell^2-2M_K^2) +R(m_\ell)
\label{approxol}
\end{eqnarray}
where the residual $R(m_\ell)$ is calculated numerically, and we  
find $R(m_e)= -0.85$ and $R(m_\mu)= -0.94$.

The normalization used in Eq.~\ref{lwct} was chosen so that 
$h(\mu)$ is naturally of order one, for example, in the 
vector-dominance model of Quigg and Jackson \cite{quijak} 
one finds that $h(M_\rho) \approx 4.3$.\footnote{The second model 
in Ref.~\cite{quijak} predicts $h(M_\rho)\approx 1.9$, the model of 
Ref.~\cite{berg} predicts $h(\mu)\approx 2.5$, and the model of 
Ref.~\cite{ampoisi} predicts $h(M_\rho)= 8 \pm 2$. Eq.~\ref{roomfora} taken 
at face value implies $h(\rho)= 8.7 \pm2$.} If we use 
the result $h(M_\rho)\approx 4.3$, assign to it an uncertainty 
from knowing the matching scale $\mu$ only to within a factor of 
two (somewhere between $M_K$ and $\Lambda \sim 1$~GeV), 
and take $\alpha=3.15$, we find 
$1.5 \lsim \hbox{Re}A \lsim 4$. Notice that this is higher than, but not 
inconsistent with, the result of Ref.~\cite{ampoisi},  
$\hbox{Re}A = 0.41 \pm 1.03$. As we will see in the next section, however, 
it is easy to construct models that take $\hbox{Re}A$ well outside of 
this range. 

It is of some interest to express the short distance 
amplitude, Eq.~\ref{kmmbu}, in this notation. Without getting into 
the issue of the relative sign between the short and long distance 
amplitudes, we can write $|h_{SD}| \approx 2.8 |\rho_0-\rho|$. 
This puts the size of the short distance contribution at the 
level of the uncertainty in the long distance contribution, making it  
clear that one cannot extract reliable information on $\rho$.

For the decay $K_L \rightarrow e^+ e^-$ the unknown combination $h(\mu)$ is 
much less important than the non-analytic terms in Eq.~\ref{approxol}.
This leads to the prediction, 
\begin{equation}
B(K_L \rightarrow e^+ e^-) = (9.0 \pm 0.5) \times 10^{-12}
\label{keebr}
\end{equation}
This result indicates that the total rate for $K_L\rightarrow e^+ e^-$ 
is expected to be two to three times larger than the absorptive 
contribution. This is in sharp contrast with the rate for 
$K_L \rightarrow \mu^+ \mu^-$ which is almost completely saturated  
by the absorptive part. This also indicates that the short distance 
contribution to $K_L \rightarrow e^+ e^-$ is completely negligible in
the standard model. We should caution the reader against taking the  
error quoted for the result, Eq.~\ref{keebr}, too seriously. It 
was obtained by fitting $h(\mu)$ from Eq.~\ref{approxol} to 
Eq.~\ref{roomfora} and using $\alpha=3.15$.

The Lagrangian of Eq.~\ref{lwct} also gives rise to the two-photon 
contribution to the decays $K \rightarrow \pi \ell^+ \ell^-$. 
In particular the decay $K^+ \rightarrow \pi^+ \mu^+ \mu^-$ has 
been studied in the literature \cite{wisa} within this 
context.\footnote{It has been pointed out in Ref.~\cite{pola} 
that it may be possible 
to use a muon polarization asymmetry in $K^+ \rightarrow \pi^+ 
\mu^+ \mu^-$ to extract information on the short distance parameter 
$\rho$. The latest results for the short distance contribution to 
such an asymmetry, Ref.~\cite{polab}, combined with the estimate 
of Ref.~\cite{wisa} for the long distance contributions indicate that 
it may be possible to extract a value for $\rho$ from a measurement 
of the asymmetry.} However, it is not possible to use that reaction to 
measure the unknown constant $h(\mu)$ (its leading contribution) 
that appears in Eq.~\ref{oneloop} 
because the counter-terms enter the amplitude for the process 
$K^+ \rightarrow \pi^+ \mu^+ \mu^-$  in the 
different linear combination\footnote{We disagree 
with the expression in Ref.~\cite{wisa} but this does not affect 
our conclusion nor does it affect the conclusions of Ref.~\cite{wisa}.} 
$h_1-2h_2+4h_3$. It is perhaps interesting to point out that 
the two-photon contribution to the reaction $K_S \rightarrow 
\pi^0 \ell^+ \ell^-$ receives contributions from Eq.~\ref{lwct} 
in the same combination as $K_L \rightarrow \ell^+ \ell^-$, 
$h_1+h_2+h_3$. Unfortunately, however, the process $K_S \rightarrow 
\pi^0 \ell^+ \ell^-$ is dominated by a one-photon intermediate 
state \cite{rafael}, and it is unlikely that one could extract the two-photon 
contribution. Moreover, the short distance Hamiltonian that contributes 
to $K_L \rightarrow \ell^+ \ell^-$ also contributes to $K_S \rightarrow 
\pi^0 \ell^+ \ell^-$ and the ratio of this short distance contribution 
to the counter-term contribution  is the same in both 
processes.

In conclusion, we find that the processes that can measure 
$h_1+h_2+h_3$ ($K_L \rightarrow \mu^+ \mu^-$, $K_L \rightarrow e^+ e^-$ and 
$K_S \rightarrow \pi^0 \ell^+ \ell^-$) cannot separate it from 
the short distance contribution to $\hbox{Re}A$. For this reason 
it is not possible to extract the CKM parameter $\rho$ from the measurement 
of $K_L \rightarrow \mu^+ \mu^-$.

\section{Beyond Chiral Perturbation Theory}

If we go beyond chiral perturbation theory, it is possible to 
calculate the dispersive contribution of the two-photon 
intermediate state to $K_L \rightarrow \ell^+ \ell^-$. This 
is accomplished by introducing a model for the form factor of 
Eq.~\ref{akgg} with good high energy behavior to make the loop 
integral finite. The price we pay is that we cannot quantify 
the dependence of the result on the assumptions of the model. 
This is what several papers in the literature 
have done in different ways \cite{quijak,berg,belgeng,ampoisi}. 
Our purpose in this section is to introduce a class of models  
that contains a few free parameters. By varying these 
parameters within a reasonable range we will show that the 
uncertainty of these model calculations is sufficiently large 
to make an extraction of $\rho$ from $K_L \rightarrow \mu^+ \mu^-$ 
impossible at present.

The behavior of the form factor $F(q_1^2,q_2^2)$ 
for values of $q_1^2,~q_2^2$ near 
the chiral symmetry breaking scale $\Lambda\sim (1~{\rm GeV})^2$ cannot 
be calculated at present. At very large momentum transfers we 
expect the form factor to take the form 
\begin{equation}
F(q_1^2,q_2^2) \rightarrow \gamma^2 {\Lambda^4 \over q_1^2   q_2^2}
\label{asymp}
\end{equation}
where $\gamma$ is another parameter that should be of order one.

There are two sources 
of uncertainty when we try to use the form factor of Eq.~\ref{expandf} 
to calculate the dispersive part of the two-photon contribution 
to $K_L \rightarrow \mu^+ \mu^-$. The first one has to do with our 
ignorance of small $q^2$ behavior, parameterized by $\alpha$, 
$\beta$ and $\beta^\prime$. Our knowledge of these numbers can 
be improved with experimental studies of the Dalitz pair conversion 
modes. The second source of uncertainty has to do with the behavior 
of the form factor for values of $q^2$ near the chiral symmetry breaking
scale. This is something that cannot be measured experimentally, and 
unfortunately, it is this region that is responsible for the 
fact that we cannot calculate the dispersive part of the two-photon 
contribution. In $\chi$PT this momentum region is responsible for 
the divergence in the loop integral, and with   
dimensional regularization we parameterize this ignorance with 
unknown counter-terms, the $h_i$ of Eq.~\ref{lwct}. The only way to 
access this information from the low energy theory is if the 
Lagrangian of Eq.~\ref{lwct} induces other processes that can be 
measured. Unfortunately this is not the case, as we argued in 
the previous section. Under these conditions we will only be able 
to make reliable predictions by solving QCD at low energies.

To see how these two sources of uncertainty enter the 
rate for $K_L \rightarrow \mu^+ \mu^-$, 
we first use a simple prescription to unitarize the form factor 
$F(q_1^2,q_2^2)$ truncated at the term proportional to $\alpha$. 
With this prescription we investigate the variation of the 
dispersive part of the $K_L \rightarrow \mu^+ \mu^-$ amplitude 
with the experimental error for $\alpha$. This model resembles 
the single pole model of Quigg and Jackson \cite{quijak} with a 
variable pole mass.

After that, we generalize our model to unitarize the expansion 
of the form factor $F(q_1^2,q_2^2)$ truncated at the term proportional
to $\beta$. After fixing $\alpha$ to its 
experimental value, our second model will contain two free parameters 
that we can use to fix $\beta$ and $\gamma$ (the constant in the 
asymptotic behavior of the form-factor). This allows us to study 
the dependence of the dispersive part of $K_L \rightarrow \mu^+ \mu^-$ 
on the low energy parameter $\beta$ and on the asymptotic parameter 
$\gamma$. Effectively this corresponds to a continuous parameterization 
of the region of $q_1^2\sim q_2^2\sim \Lambda^2$ in terms of a 
single parameter, $\gamma$, while keeping the low energy behavior 
fixed. Admittedly, this is 
only the simplest model we could find to parameterize the region 
near the chiral symmetry breaking scale, but it illustrates that 
even if the low energy behavior of $F(q_1^2,q_2^2)$ is fixed, the 
dispersive contribution to $K_L \rightarrow \mu^+ \mu^-$ is 
sufficiently uncertain that we cannot extract any conclusions about the 
short distance parameter $\rho$.

In our first model we take 
\begin{eqnarray}
F(q_1^2,q_2^2) &=& F(0,0)\biggl(1+\alpha {q_1^2+q_2^2\over \Lambda^2}
\biggr)\nonumber \\
&\approx & {F(0,0) M^4 \over (q_1^2-M^2) (q_2^2-M^2)}
\label{modelop}
\end{eqnarray}
$M = \Lambda/\sqrt{\alpha}$ can be thought of as the relevant cutoff 
mass. Using this form factor, the two-photon contribution to 
$K_L \rightarrow \ell^+ \ell^-$ is finite and can be calculated 
in terms of $M$ (or $\alpha$). It is important in this model 
to keep $M > M_K$ to avoid changing the absorptive part of the 
amplitude. The calculation is standard and closely follows 
that of Quigg and Jackson \cite{quijak}. One has an integral with five 
denominators that can be reduced to integrals with only three
denominators by using partial fractions. These integrals are done in 
terms of two Feynman parameters and it is most convenient to carry out 
the last integration (over one of the Feynman parameters) numerically. 

\begin{figure}[htb]
\centerline{\epsfxsize=2.5in\epsfbox{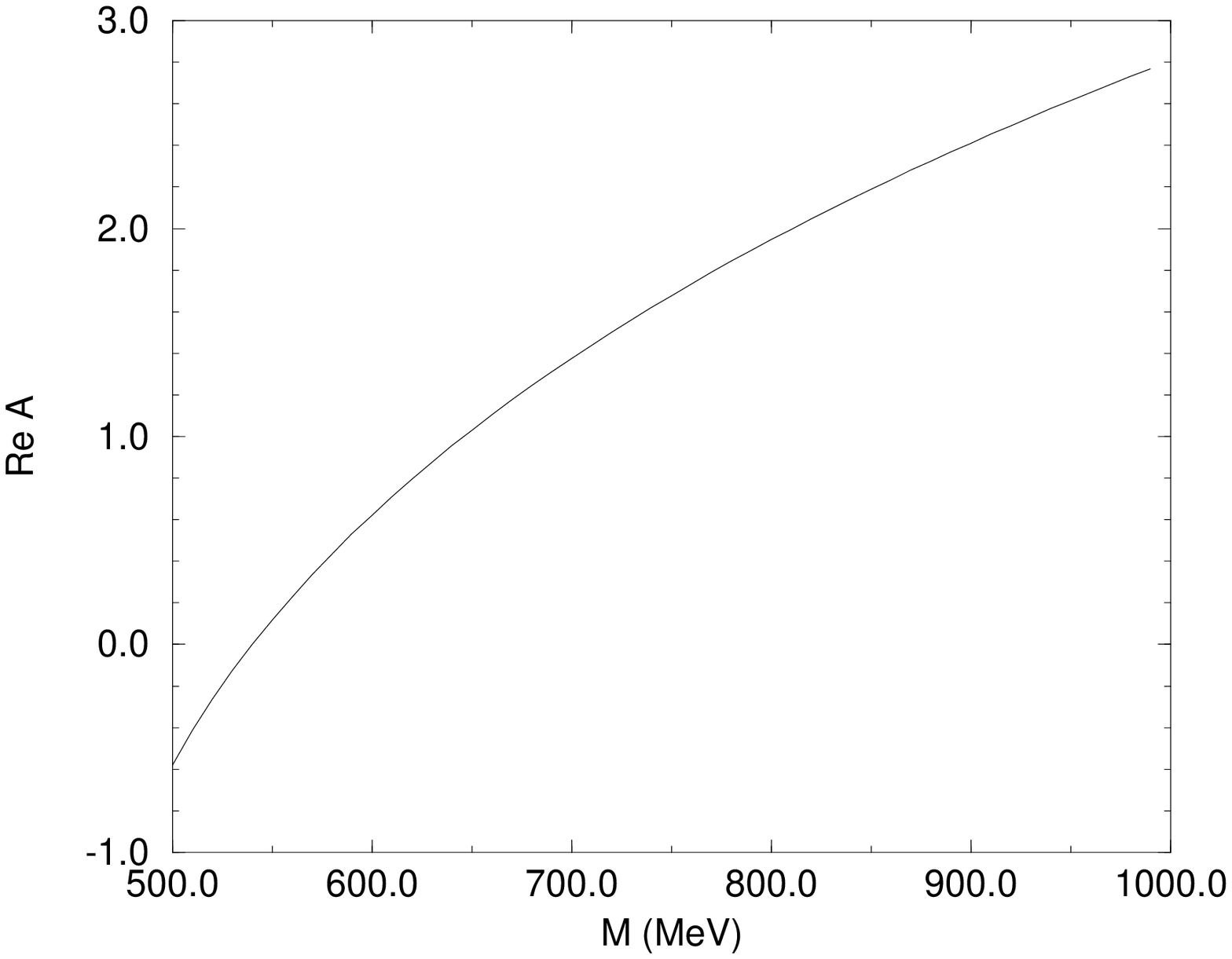}\hspace{0.2in}
\epsfxsize=2.5in\epsfbox{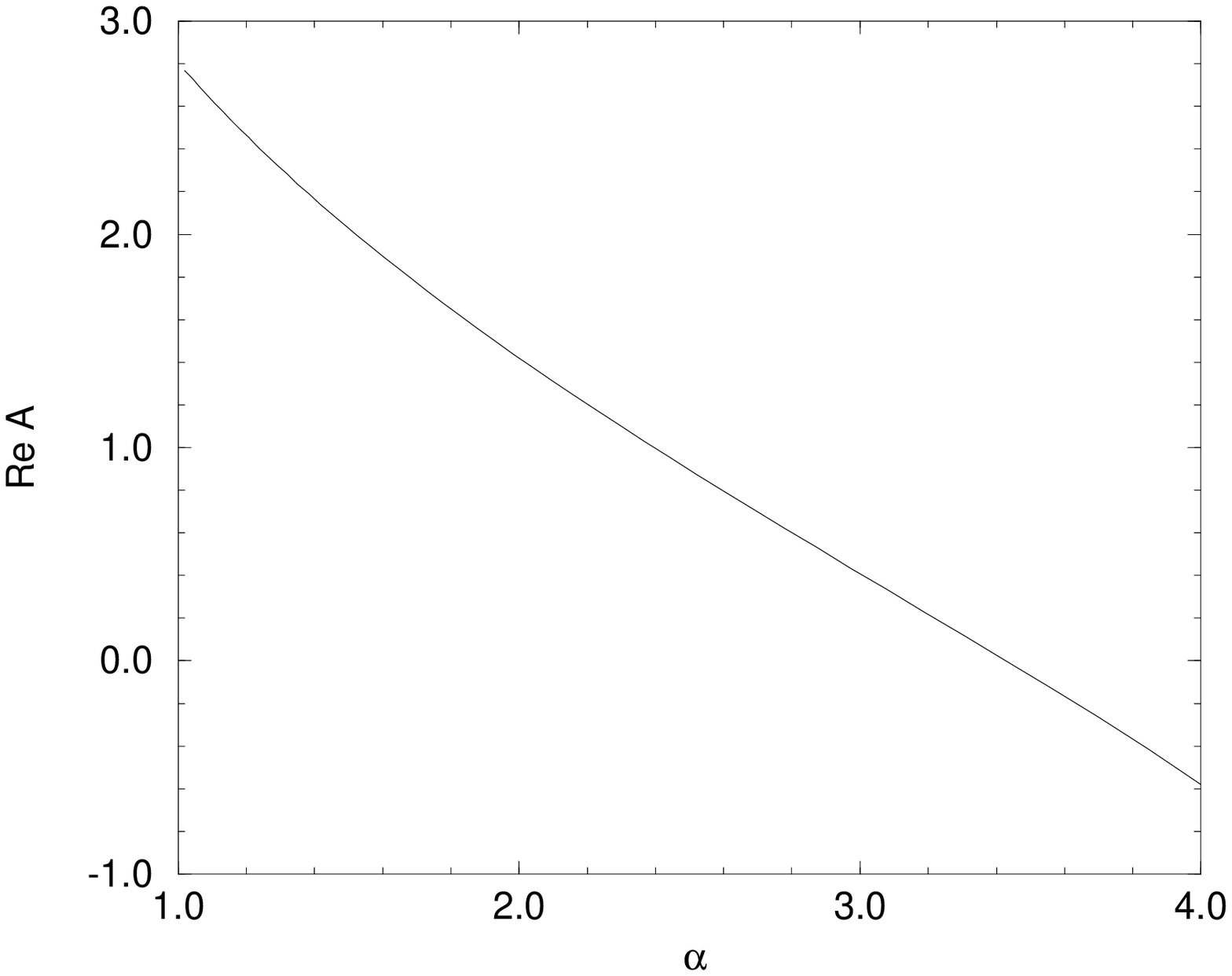}}
\caption[figure 1]{$\hbox{Re}A$ as a function of a) $M$ and b) $\alpha$ 
using the model of Eq.~\ref{amodelop}. $\alpha=3.15 \pm 0.40$ or 
$M=564 \pm 33$~MeV is the number measured in $K_L \rightarrow \ell^+
\ell^- \gamma$}
\label{f:onepole}
\end{figure}

We introduce the notation $r_\ell = m_\ell/M$ and the function
\begin{equation}
g(M_1,M_2) = \int_0^1 dx \int_0^{1-x} dy 
{M_K^2 \over M_K^2xy-M^2_1 x-M^2_2 y -m^2_\ell (1-x-y)^2}
\label{funinint}
\end{equation}
Here, the integral over $y$ can be carried out in terms of 
logarithms and inverse tangents. For our purposes, a numerical 
integral over the remaining parameter, $x$, is sufficient. With this  
model we find
\begin{eqnarray}
\hbox{Re}A &=&- {M^4 \over 2 m_\ell^2 M_K^2}\biggl[2\log r_\ell 
+ \sqrt{1-4r^2_\ell}\log\biggl({1+\sqrt{1-4r^2_\ell}\over 
1-\sqrt{1-4r^2_\ell}}\biggr)\biggr] -{1\over 2}g(0,0)
\nonumber \\
&+&\biggl({4M^2-M^2_K \over 2 M^2_K}\biggr)g(M,M)+
\biggl({M_K^2-M^2\over M^2_K}\biggr)^2g(M,0)
\label{amodelop}
\end{eqnarray}
For small values of the lepton mass it is possible to 
write the approximate analytic expression \cite{quigg}:
\begin{equation}
\hbox{Re}A \approx 
-{\pi^2 \over 12}-{1 \over 4}-\biggl(\log {m_K \over 
m_\ell }\biggr)^2+3\log \biggl({M \over m_\ell }\biggr)
\label{quiggap}
\end{equation}
The experimentally measured $\alpha = 3.15 \pm 0.40$ ($M = 564\pm
33$~MeV), corresponds in this model to $-0.13 \leq \hbox{Re}A \leq
0.62$. Notice that this result is consistent with that of Ref.~\cite{ampoisi}. 
We present in Figure~\ref{f:onepole} the resulting $\hbox{Re}A$ as 
a function of $\alpha$ and of $M$ for $M_K < M < 1$~GeV. 
It is amusing to see that $\hbox{Re}A$ crosses zero for a value 
of $\alpha$ very close to the measured $\alpha = 3.15$. For the 
corresponding value of $M$, $\hbox{Re}A$ is much smaller than 
it is for $M=M_\rho$. This is the essence of the result of 
Bergstr\"{o}m, Mass\'{o} and Singer \cite{berg}. As we will see shortly,
this result is {\it not} as much a consequence of the measured value 
of $\alpha$ (equivalently of $\alpha_{K^*}$ in \cite{berg}), as it is 
of the assumed unitarization. For this reason we regard this as an 
accidental result as we claimed in Ref.~\cite{liva}.

We now proceed to our second model, which will permit us to study 
the dependence of $\hbox{Re}A$ on the low energy parameter 
$\beta$ of Eq.~\ref{expandf} and on the parameter $\gamma$ of Eq.~\ref{asymp} 
(equivalently, on the shape of $F(q_1^2,q_2^2)$ for values of $q_i^2$ 
near $\Lambda^2$). There 
are many ways to unitarize the low energy expansion of
Eq.~\ref{expandf}. Since we do not intend to argue that there is one 
way that is better than the rest, we will simply choose a model that 
does not require any new calculations:
\begin{equation}
F(q_1^2,q_2^2) = F(0,0)
\biggl(a {M_1^2 \over q_1^2 -M_1^2}+b{M_2^2 \over q_1^2 - M_2^2}\biggr)
\biggl(a {M_1^2 \over q_2^2 -M_1^2}+b{M_2^2 \over q_2^2 - M_2^2}\biggr)
\label{modeltp}
\end{equation}
The requirement that the normalization $F(0,0)$ is not changed 
implies that $b=-(1+a)$. The requirement that the low energy 
expansion of Eq.~\ref{modeltp} matches the measured value of 
$\alpha$ gives the relation:
\begin{equation}
a={(\Lambda^2/\alpha -M_2^2)M^2_1 \over \Lambda^2/\alpha (M_2^2 
-M_1^2)}
\label{solvea}
\end{equation}
We are left with two parameters $M_1$ and $M_2$ that can be chosen 
to fix given values of $\beta$ and $\gamma$. Once again, we have the 
limitation that $M_1$ and $M_2$ must be larger than $M_K$ to avoid 
changing the absorptive part. The relation between $M_1$, $M_2$ and 
$\beta$, $\gamma$ is:
\begin{eqnarray}
{\beta \over \Lambda^4} &=& {M_1^2+M_2^2-\Lambda^2/\alpha \over 
\Lambda^2/\alpha M_1^2 M_2^2}\nonumber \\
\gamma\Lambda^2 &=& {\Lambda^2/\alpha(M_1^2+M_2^2)-M_1^2M_2^2 \over
\Lambda^2/\alpha}
\label{algamm}
\end{eqnarray}
It is easy to express the result for $\hbox{Re}A$ in terms 
of the previous model, we find:
\begin{eqnarray}
\hbox{Re}A &=& -{M_1^2 M_2^2 \over 4 m_\ell^2 M_K^2} \biggl[
2\log r_{\ell 2} + \sqrt{1-4r_{\ell 2}}\log\biggl({
1+\sqrt{1-4r_{\ell 2}}\over 1- \sqrt{1-4r_{\ell 2}}}\biggr)\biggr]
\nonumber \\ &-& {M_1^2 M_2^2 \over 4 m_\ell^2 M_K^2} \biggl[
2\log r_{\ell 1} + \sqrt{1-4r_{\ell 1}}\log\biggl({
1+\sqrt{1-4r_{\ell 1}}\over 1- \sqrt{1-4r_{\ell 1}}}\biggr)\biggr]
\nonumber \\
&-& {1\over 2}\biggl[g(0,0) + \lambda(1,M_1^2/M_K^2,M_2^2/M_K^2)
g(M_1,M_2)\nonumber \\
&-&\biggl({M_K^2-M_1^2\over M_K^2}\biggr)^2g(M_1,0)
-\biggl({M_K^2-M_2^2\over M_K^2}\biggr)^2 g(M_2,0)\biggr]
\label{amodeltp}
\end{eqnarray}
where we have now defined $r_{\ell i} = m_\ell/M_i$, and used the 
function $\lambda(x,y,z)=x^2+y^2+z^2-2(xy+xz+yz)$. In the limit 
$M_1=M_2$ Eq.~\ref{amodeltp} reduces to the result of the simpler model, 
Eq.~\ref{amodelop}.

\begin{figure}[htb]
\centerline{\epsfxsize=3.0in\epsfbox{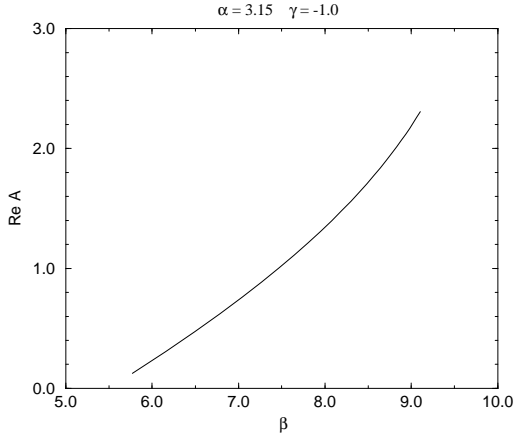}}
\caption[figure 2]{$\hbox{Re}A$ as a function of $\beta$ with 
$\alpha = 3.15$ and $\gamma = -1.0$ using the 
model of Eq.~\ref{modeltp}.}
\label{f:twopoleb}
\end{figure}

We first use this model to test the dependence of $\hbox{Re}A$ on 
the low energy constant $\beta$. For this we fix $\alpha = 3.15$ and 
$\gamma = -1.0$ and allow $\beta$ to cover the allowed range 
(determined by the requirement that both $M_{1,2}> M_K$). For 
this choice of $\gamma$ one of the mass scales is between $M_K$ and 
1~GeV, whereas the second one is between 1-3~GeV. We present 
this result in Figure~\ref{f:twopoleb}. 
Since all the points in Figure~\ref{f:twopoleb} correspond 
to $\alpha=3.15$, and have the same $\gamma$, this figure shows 
the variation of $\hbox{Re}A$ with the shape of the form 
factor at energies that can be probed experimentally. 

\begin{figure}[htb]
\centerline{\epsfxsize=3.0in\epsfbox{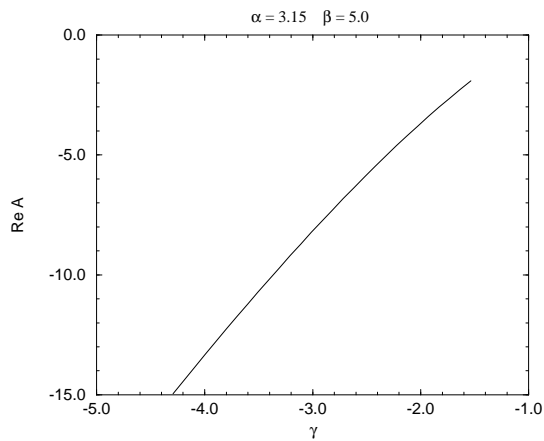}}
\caption[figure 3]{$\hbox{Re}A$ as a function of $\gamma$ with 
$\alpha = 3.15$ and $\beta=5.0$ using the 
model of Eq.~\ref{modeltp}.}
\label{f:twopoleg}
\end{figure}

We next fix $\beta$ (as we would do once this parameter is measured 
in $K_L \rightarrow \ell^+ \ell^- \gamma$) and vary $\gamma$ to 
see the effect of varying the shape of the form factor in the 
region $q_i^2 \sim \Lambda^2$. For definiteness we pick $\beta=5$, 
$\alpha = 3.15$ and allow $\gamma$ to vary in the acceptable region 
(with $M_{1,2} >M_K$). We show this result in Figure~\ref{f:twopoleg}.
This figure shows that the smallness of $\hbox{Re}A$ 
found by Bergstr\"{o}m, Mass\'{o} and Singer \cite{berg} is 
{\it not} a consequence of the experimental value $\alpha=3.15$, but 
instead it is a consequence of the specific unitarization of the form factor 
chosen by those authors. This figure also emphasizes the fact 
that measurements of the low energy parameters $\alpha$, $\beta$ 
and $\beta^\prime$ will not remove the uncertainty in the 
calculation.

\begin{figure}[htb]
\centerline{\epsfxsize=2.0in\epsfbox{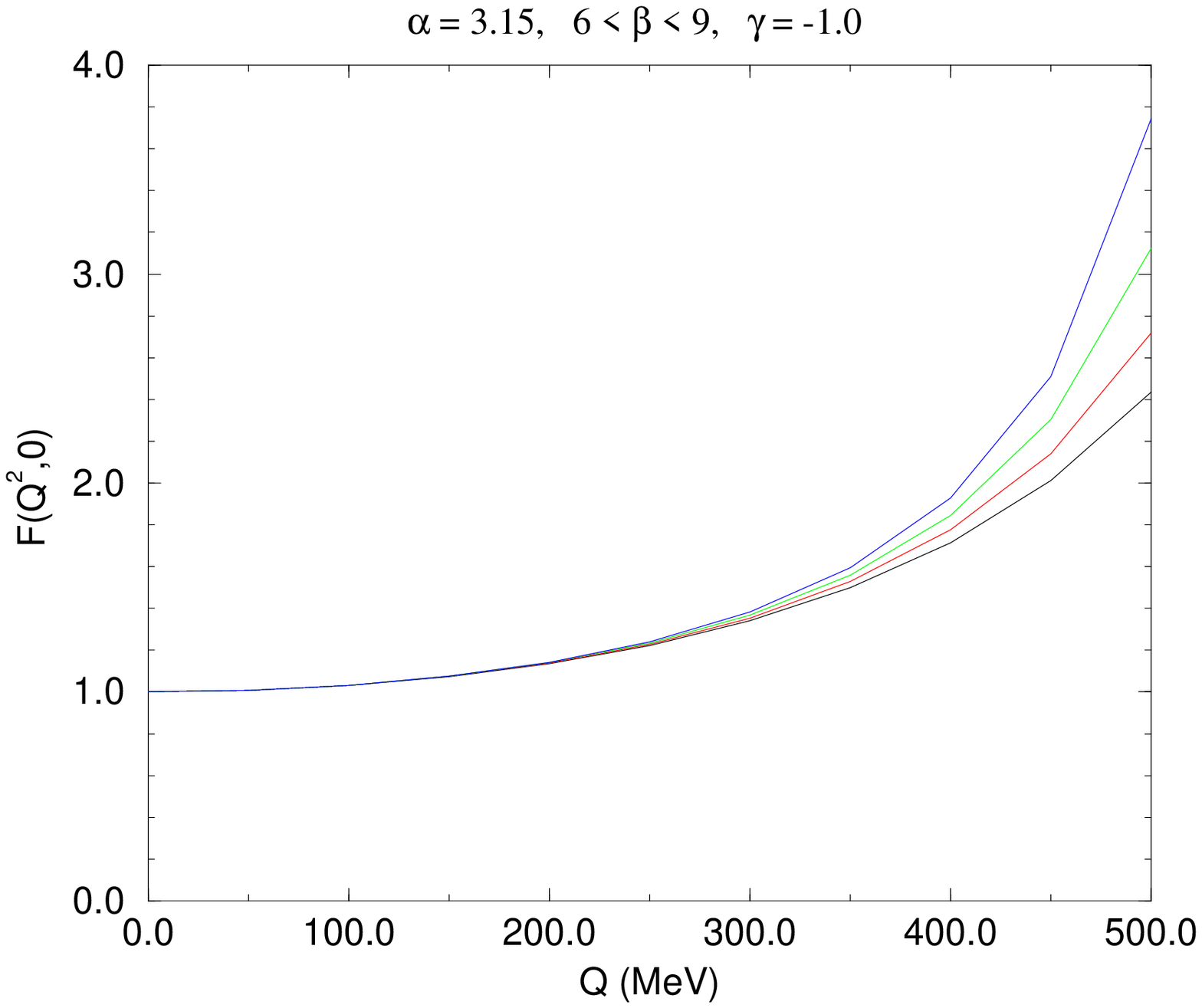}\hspace{0.2in}
\epsfxsize=2.0in\epsfbox{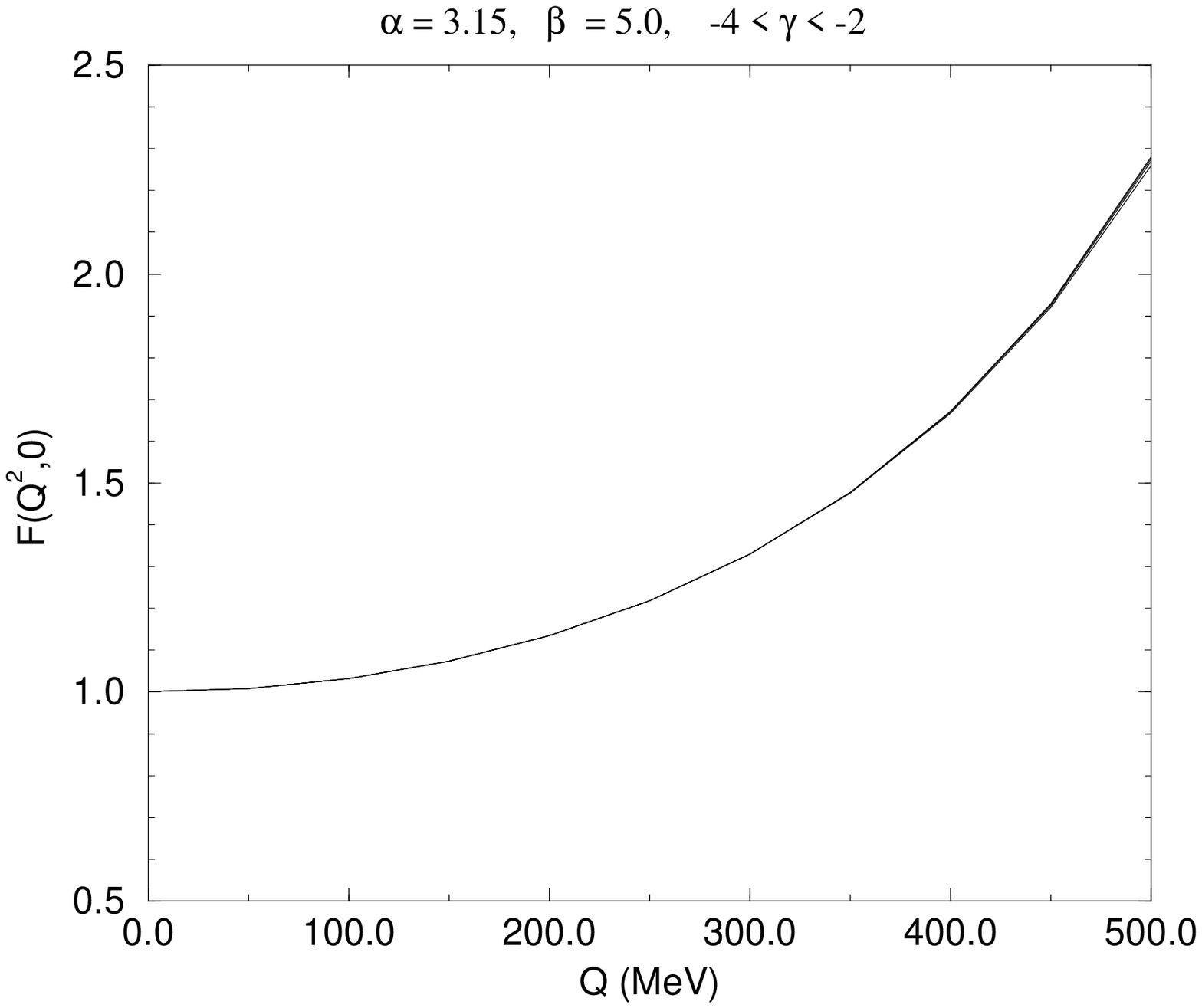}}
\caption[figure 4]{$F(q^2,0)$ for values of $\alpha$, $\beta$ and 
$\gamma$ used in Figures~\ref{f:twopoleb}~and~\ref{f:twopoleg}.}
\label{f:formfactor}
\end{figure}

Finally, in Figure~\ref{f:formfactor} we plot $F(q^2,0)$ for 
values of $\beta$ and $\gamma$ used in 
Figures~\ref{f:twopoleb}~and~\ref{f:twopoleg} in the region 
of $q^2$ that can be probed in $K_L \rightarrow \ell^+ \ell^- \gamma$.  
This figure shows that the variation in the form-factor responsible 
for the large allowed range of $\hbox{Re}A$ of Figure~\ref{f:twopoleg} 
cannot be seen at all in the region probed by 
$K_L \rightarrow \ell^+ \ell^- \gamma$. This illustrates our 
claim that a better measurement of  $K_L \rightarrow \ell^+ \ell^-
\gamma$ will not remove the uncertainty in $K_L \rightarrow \mu^+ \mu^-$.

Before ending this section we should compare our model calculation 
to that of Ref.~\cite{ampoisi}. It may be argued that the model 
of Ref.~\cite{ampoisi} has a better physical motivation than the 
toy models that we use. The authors of Ref.~\cite{ampoisi} construct 
their model to incorporate three ingredients: the known low energy 
behavior of $F(q_1^2,q_2^2)$ (our $\alpha$); the fact that the only 
poles in the region between the kaon mass and 1~GeV are those generated 
by the vector meson resonances; and the high $q^2$ behavior from 
perturbative QCD. From these ingredients they reach the conclusion 
that the uncertainty in $\hbox{Re}A$ is under some control and 
obtain their bound on $\rho$. Our differences arise because we argue that 
vector meson dominance for weak decays is nothing more than an 
assumption. Our toy models allow us to vary the behavior of 
$F(q_1^2,q_2^2)$ in the region near one GeV, where neither chiral 
perturbation theory nor perturbative QCD are reliable. We are thus 
able to show that $\hbox{Re}A$ is quite sensitive to the behavior 
of the form factor in this matching region, and we argue that this 
is, in fact, the main source of uncertainty. This is why we conclude 
that it is impossible at present to estimate $\hbox{Re}A$ reliably. 
We do not claim to have a physically motivated model for the behavior 
of $F(q_1^2,q_2^2)$ in the matching region, but simply point out that 
the conclusion of Ref.~\cite{ampoisi} follows from their assumption 
about the behavior of $F(q_1^2,q_2^2)$ in the matching region. We 
believe that it will not be possible to test that assumption with 
any precision in the foreseeable future, and, therefore, disagree with 
their conclusion. 

\section{Conclusions}

We have studied the dispersive contribution of the two-photon 
intermediate state to the decay $K_L \rightarrow \mu^+ \mu^-$. 
Within chiral perturbation theory we can write the real part 
of the amplitude in terms of one unknown combination  $h(\mu)$. 
The current experimental measurements of $K_L \rightarrow \gamma 
\gamma$ and $K_L \rightarrow \mu^+ \mu^-$ imply that $h(\mu)=8.7 \pm 2$. 
If we cast the short distance amplitude 
in the same notation, we find that $h_{SD} \sim 3$, of the same 
order as the uncertainty in the long distance dispersive contribution. 
We find that there is no process where $h(\mu)$ can be determined 
separately from $h_{SD}$ and, therefore, conclude that it is 
impossible to extract information on short distance parameters 
from the measurement of $K_L \rightarrow \mu^+ \mu^-$ 
within chiral perturbation theory.

We have also found that the rate for $K_L \rightarrow e^+ e^-$ 
is relatively insensitive to the precise value of $h(\mu)$. From 
this observation we are able to predict $B(K_L \rightarrow e^+ e^-)
\sim 9 \times 10^{-12}$. 

We have also re-visited the calculation of $\hbox{Re}A$ within 
models. We have used simple pole models to unitarize the high 
energy behavior of the off-shell $K\gamma^* \gamma^*$ vertex as is 
common in the literature. We have used a class of models that 
can accommodate the measured low-energy behavior of said vertex 
while allowing us to vary the form factor in the region near the 
chiral symmetry breaking scale. From this exercise we find that, the main 
source of uncertainty in $\hbox{Re}A$ {\it does not} arise from 
the error in the measurement of $\alpha$. In fact, we find that 
even if the low energy behavior of the $K\gamma^* \gamma^*$ vertex 
is precisely measured (in our parameterization this means that 
$\alpha$ and $\beta$ are well measured), the uncertainty arising 
from the energy region that is not accessible to experiment is 
sufficiently large to make $\hbox{Re}A$ impossible to predict. 

Our conclusion is, therefore, rather pessimistic. We believe that 
it will be impossible to extract any information on the short 
distance parameter $\rho$ from a measurement of $K_L \rightarrow 
\mu^+ \mu^-$ in the foreseeable future. This situation will change 
only when we are able to calculate reliably the long distance 
amplitude from QCD.

\vspace{0.5in}
\noindent{\bf Acknowledgments}

\noindent This work was supported in part 
by the DOE OJI program under contract number DE-FG02-92ER40730. 
I am grateful to the theory group at SLAC for their hospitality 
while part of this work was performed. I wish to thank G.~Isidori 
and G.~d'Ambrosio for useful discussions
and for a critical reading of this manuscript. I also thank  
J.~Donoghue, L.~Littenberg and C.~Quigg for comments on the manuscript.

\end{document}